\documentclass[12pt,preprint]{aastex}
\usepackage{graphicx}
\usepackage{longtable}
\usepackage{multirow}
\usepackage{url}
\usepackage{lipsum}

\begin{document}

\title{On the inconsistency between cosmic stellar mass density and star formation rate up to $z\sim8$}

\author{H. Yu\altaffilmark{1} and F. Y. Wang\altaffilmark{1,2}{*}}
\affil{
$^1$ School of Astronomy and Space Science, Nanjing University, Nanjing 210093, China\\
$^2$ Key Laboratory of Modern Astronomy and Astrophysics (Nanjing University), Ministry of Education, Nanjing 210093, China\\
} \email{{*}fayinwang@nju.edu.cn}

\begin{abstract}
In this paper, we test the discrepancy between the stellar mass
density and instantaneous star formation rate in redshift range
$0<z<8$ using a large observational data sample. We first
compile the measurements of the stellar mass densities up to $z\sim
8$. Comparing the observed stellar mass densities with the
time-integral of instantaneous star formation history, we find that
the observed stellar mass densities are lower than that implied from
star formation history at $z<4$. We also use Markov chain
monte carlo method to derive the best-fitting star formation history
from the observed stellar mass density data. At $0.5<z<6$, the
observed star formation rate densities are larger than the best-fitting one,
especially at $z\sim2$ where by a factor of about two. However, at lower ($z<0.5$) and higher
redshifts ($z>6$), the derived star formation history is consistent
with the observations. This is the first time to test the
discrepancy between the observed stellar mass density and
instantaneous star formation rate up to very high redshift
$z\approx8$ using the Markov chain monte carlo method and a varying
recycling factor. Several possible reasons for this discrepancy are
discussed, such as underestimation of stellar mass density, initial
mass function and cosmic metallicity evolution.

\end{abstract}


\section{Introduction}
Thanks to the development of the telescopes, more and more important
and accurate data are obtained in almost all fields of astronomy.
The multi-wavelength observations with Hubble, Spitzer Space
Telescopes and other large ground-base telescopes give us much
information of high redshift galaxies. It gives us a great chance to
measure the stellar mass density (SMD) from galaxy surveys
\citep[hereafter G15]{Labbe2013,Santini2015,Grazian2015} and instantaneous star
formation rate density (SFRD)
\citep{Bouwens2012a,Bouwens2012b,Schenker2013} up to about $z=8$. Because of the physical connection between the star formation history (SFH) and
SMD, we expect that these two measured quantities should be
consistent with each other.

Integrating the instantaneous SFH
over redshift and making some correction for the mass-loss
during the stellar evolution through stellar winds and explosion
processes \citep{RenziniVoli1981,WoosleyWeaver1995}, we can get the
predicted stellar mass density history (SMH). Meanwhile,
after the same stellar evolution correction, the SFH can be
obtained from derivative of SMH by redshift. Just as mentioned
above, both of the SFH and SMH can be measured independently. So one can
compare them to check whether the instantaneous SFH and SMH are
consistent with each other. However, in order to make a
reasonable comparison, the SFH and SMH should be derived under same
assumptions, such as initial mass function (IMF), metallicity and dust
correction. Furthermore, to get the unbiased SFH and SMH, we also
need to obtain the correct SFRD and SMD calibration respectively.
Because all of these factors can affect the final comparison result
\citep[G15;][]{HopkinsBeacom2006,Wilkins2008,MadauDickinson2014}.

Numerous studies have been done to compare those two quantities, but
their results are not consistent with each other. For example, some
authors have found that there is a good agreement between the
measured SFH and SMH
\citep{MadauPozzetti1998,Fontana2004,Arnouts2007,Reddy2011,Behroozi2013}.
\cite{Reddy2011} found that the SFH and SMH had a general
agreement with each other if considering some systematic effects,
such as the evolution of the UV luminosity function of galaxy,
stellar mass in UV faint galaxies and the dust attenuation varying
with luminosity. \cite{Behroozi2013} constrained the SFH and SMH
based on the stellar mass-halo mass relation and found that both SMH
and SFH are consistent with observations. On the contrary, other
authors have found that there is puzzling disagreement between them
\citep[G15;][]{HopkinsBeacom2006,Wilkins2008,Santini2012,MadauDickinson2014}.
\cite{HopkinsBeacom2006} (hereafter, HB06) found that the
SMH inferred from the best-fitting SFH is larger than the observed
one over redshift $z<3$, and the larger factor peaked at redshift
$z\sim2$ is about four. \cite{Wilkins2008} (hereafter, W08) compared
the SMD and SFRD in redshift $z\leq4$. They found the derived SFH
from observed SMH is lower than the observed SFH and the discrepancy
peaks at $z\sim3$ by 0.6 dex. \cite{MadauDickinson2014} (hereafter,
MD14) have also found this discrepancy, although it is not so
significant. They have found that the inferred SMD history
was larger than the observed one by a factor about only 60\%.
However, these work used different stellar IMF assumptions. HB06
used the IMFs, noted as BG IMF and SalA IMF in HB06, given in
\cite{Baldry2003}. W08 defined a new IMF with low-mass slope of -1.0
($0.1\,M_\odot<M<0.5\,M_\odot$), and used this IMF in their work.
While MD14 used a traditional Salpeter IMF with slope of -2.35. The
choices of IMF can lead to some deviations in the comparison, which
we will show in the discussion section. Several explanations are
proposed for this discrepancy, including underestimation of the
stellar mass \citep{Maraston2010,Bernardi2013,Courteau2014}.
Instead, some authors claimed that the observed SFRD is
overestimated in UV and IR band \citep{Utomo2014} or FUV and U band
\citep{Boquien2014}. Alternatively, a possible evolution of
stellar IMF will also affect the estimations of observed
SFH and SMH (HB06; W08). Because this discrepancy is still under
debate, we use a large observed SFRD and SMD data, including
the latest observations, to reinvestigate this problem by using
Markov chain monte carlo (MCMC) method and considering the stellar
evolution effect in a more detailed way.

The structure of this paper is organized as follows. We will give an
introduction of the SMD and SFH in section 2 and 3 respectively. In
section 4, we will introduce our method and give our result.
Some potential causes for the discrepancy between observed
SFH and SMH are discussed in section 5. Finally, summary will be
given in section 6. For simplicity, we assume the Salpeter IMF with
index of -2.35 \citep{Salpeter1955} and solar metallicity as a
universal metallicity in our work. The flat $\Lambda$CDM cosmology
with $h=0.7$ and $\Omega_m = 0.3$ is adopted.

\section{The Stellar Mass Density}
The SMD $\rho_*(z)$ is the stellar mass in a unit co-moving cosmic
volume at redshift $z$. It can be obtained by integrating the galaxy
stellar mass function (GSMF) $\Psi(M)$ at a certain redshift,
\begin{equation}\label{gsmf}
  \rho_*(z)=\int^{\infty}_0 M\Psi_z(M)~dM,
\end{equation}
where $\Psi(M)~dM$ represents the number of galaxies with mass
between $M$ and $M+dM$ in a unit co-moving cosmic volume. In
practice, we integrate the GSMF from $M_{min}$ to $M_{max}$ instead
of $0$ to $\infty$, where the $M_{min}$ and $M_{max}$ are the low
and high limits of stellar mass of galaxies. Generally, the
$M_{min}$ and $M_{max}$ are taken as $10^8\,M_\odot$ and
$10^{13}\,M_\odot$, respectively. In this work, all the SMD data
from G15 are obtained from integrating their GSMF over range of
$10^8M_\odot<M<10^{13}M_\odot$. However MD14 adopted range about
$10^{9.5}M_\odot<M<10^{13}M_\odot$ (see MD14 for more detail).
Fortunately, this little difference of low-mass limit doesn't make
much difference to the observed SMD (MD14). Carefully, we also
compare those observed SMD data in G15 and MD14 and find that the
difference is less than 1\% which can be neglected.

Fundamentally, the method of estimating the SMDs is to fit the
observed galaxy spectral energy distributions (SEDs) with a library
of template SEDs. Then we can obtain the optimal mass-to-light ratio
$M/L$ parameter of the galaxies. It should be noted that the IMF
plays a very important role in estimating the galaxy stellar mass.
It represents the number ratio of stars with a certain mass among a
stellar population which includes all of stars formed at the same
time. Usually, the bright massive stars emit almost all the light of
a galaxy while the faint low-mass stars dominate the stellar mass of
a galaxy so the low-mass slope of the IMF affects the estimating of SMD.
Meanwhile, there is a remarkable difference in the
evolution of stars with different masses. The massive stars evolve
faster and loss more mass than low-mass stars. Therefore,
assumptions on IMF will affect the mass-to-light ratio $M/L$ of a
galaxy as well as the recycling fraction of stellar mass, which
means the mass fraction of each generation of stars return into the
interstellar medium through stellar wind, explosion or some other
processes. All of these will bias the estimation of the galaxy
stellar mass. For simplicity, a simple power-law IMF of
\cite{Salpeter1955} in the mass range $0.1~M_{\odot}-100~M_{\odot}$
is adopted in our work, although it is challenged by some
observations. There are also some other IMFs used in previous works
\citep{Bastian2010}, such as Chabrier IMF \citep{Chabrier2003} and
Modified Salpeter IMF \citep{Baldry2003}. The converting
factor of SMD from one IMF to another can be obtained using the
population synthesis code like PEGASE \citep{FiocRocca1997} \textbf{or FSPS \citep{Conroy2010}.}

Metallicity is another important factor to affect the estimating of SMDs. Low-matallicity star evolves faster while the
high-metallicity star will loss more mass since the strong stellar
wind. Therefore, different metallicities will give different
recycling fraction of stellar mass. Moreover since the
average cosmic matallicity evolves with redshift or the age of the
univese, we should, in principle, use different matallicities at different
redshifts. But it is so complex that we just adopt the solar
metallicity $Z_{\odot}=0.02$ as the universal metallicity in our
work, which is consistent with that of MD14. We will leave the systematic bias analysis of metallicity assumption in discussion section.

Thanks to the large galaxy surveys such as SDSS and 6dFGRS, and also
some large telescopes such as HST, Spitzer and VLT, SMD data can be
accurately measured to $z\sim8$. We choose 124 observed SMD
data from previous literature over radshift range $0<z\le8$. Since the
SMD data from different groups might be estimated in different IMFs,
we should rescale them to Salpeter IMF. Luckily, these SMD data
scaled by Salpeter IMF can be found in MD14 and G15. All of these
data and their references are listed in Table 2.

\section{The Star Formation History}
Determining the cosmic SFH is a key problem in many fields of
astronomy, such as the formation of galaxy and the cosmic
metallicity evolution. Many works have been done to measure the
cosmic SFH using different methods. HB06 used UV and IR
luminosities as the tracers to measure the SFRD up to $z\approx6$.
There are also other tracers of the SFRD such as the H$\alpha$ line,
radio and X-ray emissions (for a review, see MD14). SFRs are
usually measured from the typical information of very massive stars,
since they have very short life compared with the typical star
forming timescale. The UV emission of a newly formed stellar
population is dominated by those massive stars, so it can be an
instantaneous indicator of the SFRs
\citep{Kennicutt1998,Salim2007,HaardtMadau2012,Schenker2013}.
Besides, since the interstellar dust can absorb the UV emission from
those massive stars and re-radiate at MIR and FIR wavelengths, the
IR observations can be another important indicator of the SFRs
\citep{Magnelli2011,Magnelli2013,Gruppioni2013}. Generally,
the dust extinction at FIR band is negligible while at UV band is
much more significant, especially for those star forming regions
surrounded by dense clouds. Therefore, the correction of dust
extinction for using the UV luminosity as SFRD indicator is very
important. For the IR band radiation, since the dust can be also
heated by old low-mass stars or AGNs, it is not good enough to use
IR luminosity to estimate the SFRD when the cosmic SFR is very small
or for those galaxies with low SFR such as our Milky Way
\citep{Lonsdale1987}. The IR luminosity becomes a robust tracer to
SFR at $1<z<4$, where the larger SFR makes the new born massive
stars dominating the dust heating. However, the IR detector is not
sensitive enough to measure the IR luminosity of high-redshift
galaxies, while the UV emission can be measured easier at $z>1$ as
it is redshifted to optical band. Therefore, combining UV and IR
observations can give us a better estimation of the cosmic SFRD over
all redshift range. What's more, since the short life of the
massive star, the death events of massive stars can also be tools to
measure the SFRs, such as core-collapse supernovae
\citep{Dahlen2004,Li2011,Horiuchi2013} and long duration gamma-ray
bursts \citep{WangDai2009,Kistler2009,WandermanPiran2010}.

Being same as SMD, the estimation of SFRD also depends on the choice
of IMF since those indicators can only trace the formation rates of
massive stars. We need to factor the total SFR over the entire mass
range based on an IMF assumption. MD14 chose the SFR data estimated
from FUV and IR data based on the Salpeter IMF and gave the SFH up
to $z\sim8$ by fitting the observed data. It has form as
\begin{equation}\label{eq_MadauSFH}
\psi(z)=a\frac{(1+z)^b}{1+[(1+z)/c]^d}\rm~M_{\odot}~yr^{-1}~Mpc^{-3}
\end{equation}
with the optimal parameters $(a,b,c,d)=(0.015,2.7,2.9,5.6)$.
\cite{Cole2001} also gave another form of SFH as
\begin{equation}\label{ep_ColeSFH}
\psi(z)=\frac{h(a+bz)}{1+(z/c)^d},
\end{equation}
where the $h=0.7$ is reduced Hubble constant. The optimal parameters
are $(a,b,c,d)=(0.0166,0.1848,1.9474,2.6316)$ after considering the
dust extinction. The data without correction for absorption yields
$(a,b,c,d)=(0.0,0.0798,1.658,3.105)$. HB06 used a modified Salpeter
IMF, which noted as SalA IMF, and gave the optimal parameters as
$(a,b,c,d)=(0.0170,0.13,3.3,5.3)$. In following analysis, we will
use both of these two forms of SFH to remove the possible effect of
SFH forms. For the form in MD14, we simply adopt their optimal
parameters. While for the Cole form, we adopt the optimal parameters
gave in HB06. Because of the different choices of IMF, we use the
factor 0.77 suggested in HB06 to convert the SFH to Salpeter one.

\section{Methods and Results}
The cosmic SMD at a certain redshift is the cumulative mass of all
the stars formed at higher redshifts. Therefore the SMD $\rho_*(z)$
can be expressed by the integration of the SFH $\psi(z)$ as (MD14)
\begin{equation}\label{relation_sfr_smd_1}
  \rho_*(z)=(1-R)\int_0^{t(z)}\psi(t^{\prime})dt^\prime=(1-R)\int_z^{\infty}\psi(z^\prime)\frac{dz^\prime}{H(z^\prime)(1+z^\prime)},
\end{equation}
where $H(z)=H_0\sqrt{\Omega_m(1+z)^3+(1-\Omega_m)}$ is the Hubble
parameter in a flat $\Lambda$CDM cosmology, and the recycling
fraction factor $R$ represents the mass fraction of each stellar
population returned to the interstellar medium. This fraction factor
can be obtained by using stellar population synthesis code. In
previous work, MD14 used a constant fraction factor $R=0.27$ while
G15 used a value of 0.28.

The above equation is just an approximation of the actual mass
recycling process because it is based on an assumption that the
recycling process happens instantaneously. However, a newly stellar
generation would have returned only little mass while a generation
formed at early time would have returned more mass into interstellar
medium. Therefore, a more accurate equation should be expressed as
(W08)
\begin{equation}\label{relation_sfr_smd_2}
  \rho_*(z)=\int_0^{t(z)}\psi(t^\prime)(1-f_r(t-t^\prime))dt^\prime,
\end{equation}
where $f_r(t-t^\prime)$ is the mass fraction of the stellar
generation, which formed at $t^\prime$, have been returned into
interstellar medium at time $t$. If $f_r(t-t^\prime)$ is a constant,
equation (\ref{relation_sfr_smd_2}) reduces to equation
(\ref{relation_sfr_smd_1}).

\begin{figure}[hbt]
  \centering
  \includegraphics[width=10cm]{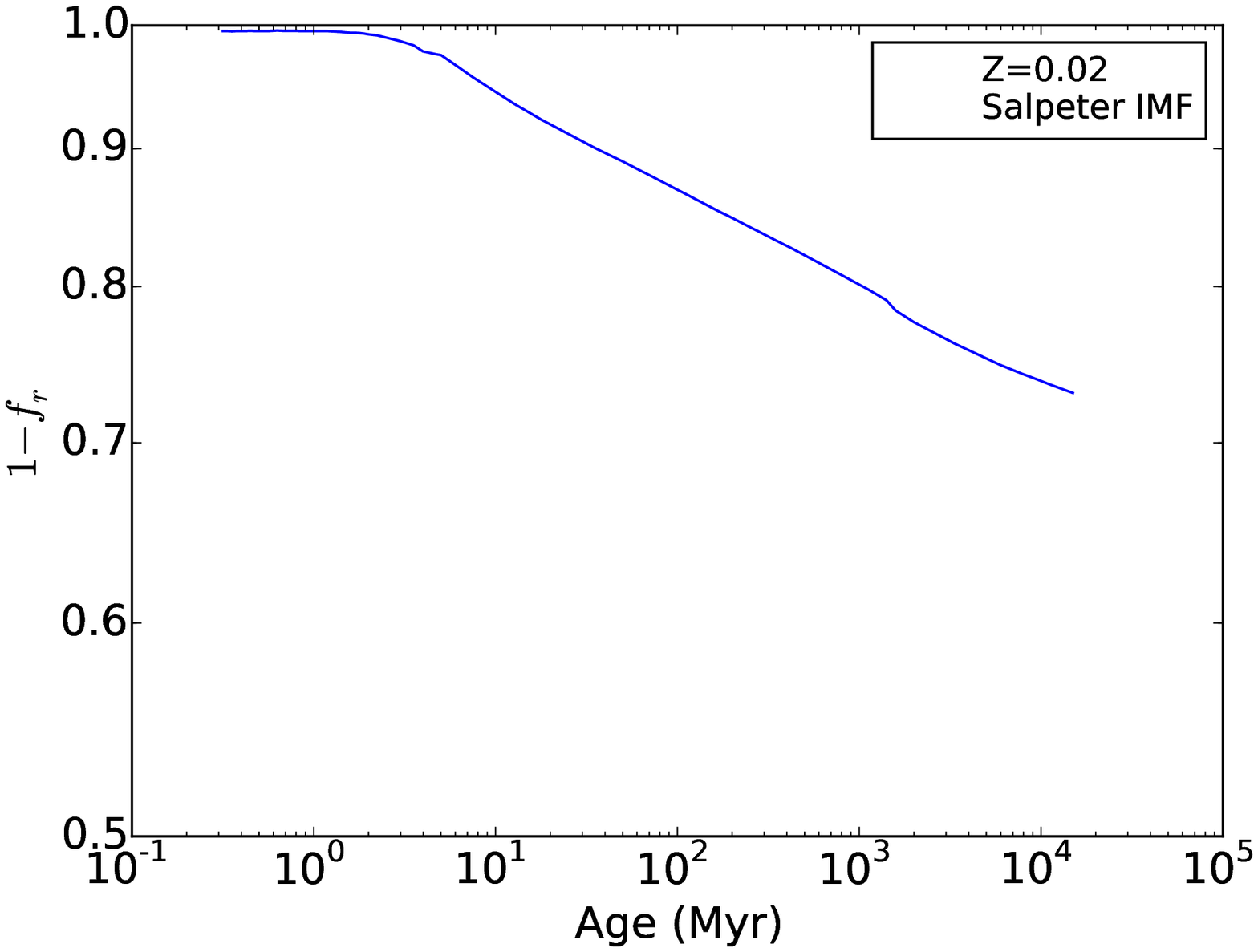}\\
  \caption{The mass evolution of a stellar population formed at same time with its age. The $f_r$ is ``recycling fraction" factor.}\label{ftfig}
\end{figure}

To calculate $f_r(t-t^\prime)$, we consider the mass evolution of an
instantaneously formed stellar population using the FSPS code
\citep{Conroy2010}. This code can give the evolution of the current
remained stellar mass fraction, which means $1-f_r$, of a simple
stellar population after setting some necessary parameters. We
choose the Salpeter IMF and solar metallicity while leaving other
parameters as their default values in the code\footnote{For the
parameters in the code, we use $\rm verbose=0$, which means the
Padova isochrones model is used. The parameter of the dust
absorption model is $\rm dust\_type=0$, which is corresponding to
the power-law attenuation model. For the dust emission model, $\rm
add\_dust\_emission=1$ means the Draine \& Li (2007) model. We
consider the nebular continuum component, i.e., $\rm
add\_neb\_continuum=1$. More detailed information can be found in
the manual for FSPS 2.5 code, which can be downloaded on
github.com/cconroy20/fsps.}. Fig. \ref{ftfig} shows the evolution of
the current mass fraction $1-f_r(t)$ of a simple stellar population.
We can see that it has almost no mass loss within 1 Myr, then the
fraction is up to about 0.27 at $14$ Gyr. If we just adopt a
constant recycling fraction factor $R=0.27$, we will over-estimate
the mass-loss effect in stellar evolution. The choice of IMF and
metallicity will affect the evolution of $f_r$, which will be
discussed in detail in discussion section.

Given the form of SFH $\psi(z)$, we can predict the SMH with
equation (\ref{relation_sfr_smd_1}) or equation
(\ref{relation_sfr_smd_2}). Both MD14 and G15 used equation
(\ref{relation_sfr_smd_1}), but their recycling fraction factors are
slightly different, which are 0.27 and 0.28 respectively. In our
work, we use the evolving $f_r$ instead of the constant one
(hereafter, $f_r$ represents the evolving recycling factor
while the $R$ represents the constant one). Since many previous
works have been done to determine the evolution of cosmic SFH, we
can use their results to predict the SMH with our $f_r$.

\begin{figure}[hbt]
  \centering
  \includegraphics[width=10cm]{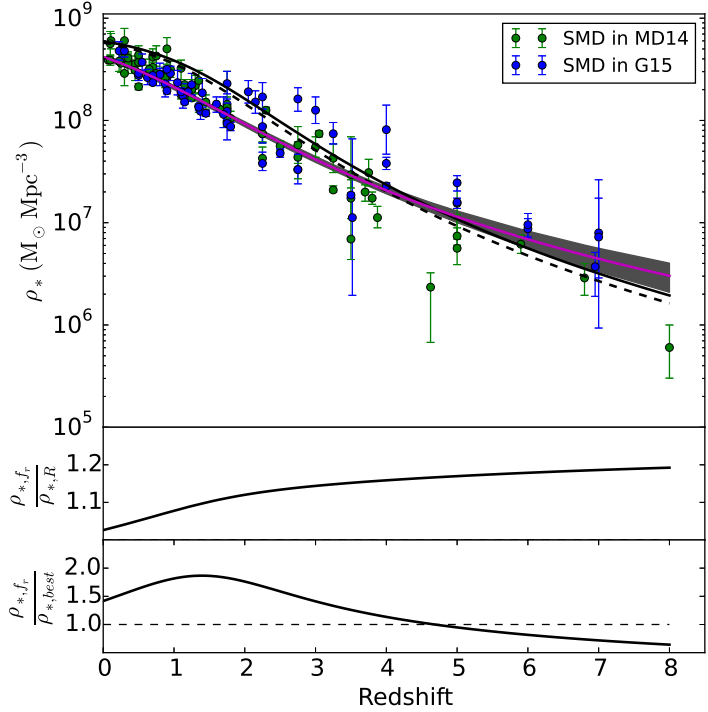}\\
  \caption{The redshift evolution of SMH.
   We use the SFH form of MD14 and their best-fitting parameters in this case. Top panel: the green
   and blue circles with $1\sigma$ errors are the observed SMD data given in MD14 and G15,
   respectively. The black solid line represents the SMH predicted from observed SFHs
   with an evolving factor $f_r$ and the black dot-dashed line represents the SMH predicted
   from observed SFHs with a constant factor $R$, respectively.
   The magenta line and the gray region are the inferred SMH from our best-fitting SFH from the observed SMD data and $95\%$
   confidence region obtained with MCMC method. Middle panel: the ratio of the predicted SMHs
   from observed SFHs between evolving $f_r$ and constant $R$ at different redshift.
   It is up to about $1.2$ at $z=8$. Bottom panel: the ratio between SMHs from observed SFHs in MD14
   and our best-fitting SFHs from observed SMD data with evolving $f_r$. It peaks at $z\sim1.5$ with factor about 2.}\label{SMDfig1}
\end{figure}

\begin{figure}[hbt]
  \centering
  \includegraphics[width=10cm]{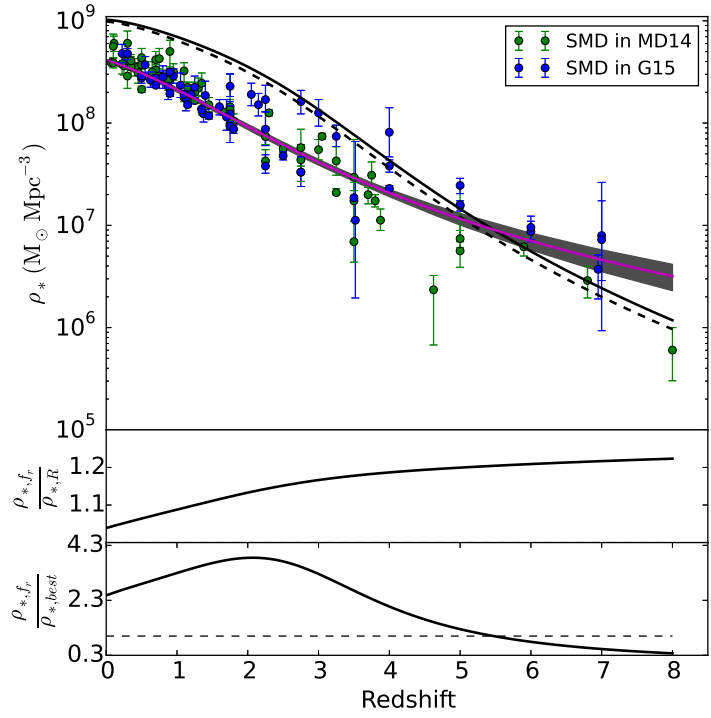}\\
  \caption{Same as Fig. \ref{SMDfig1} but using the SFH form of \cite{Cole2001}
  and observed SFH given in HB06. In the bottom panel, the ratio peaks at $z\sim2$ with factor about 4. }\label{SMDfig2}
\end{figure}

In the top panel of Fig. \ref{SMDfig1}, the green and blue
solid circles with $1\sigma$ errors are the observed SMD data given
in MD14 and G15. The black solid line represents the SMH predicted
from observed SFH in MD14 with $f_r$ while black dot-dashed line
represents the one derived with $R=0.27$. The top panel of Fig.
\ref{SMDfig2} is same as Fig. \ref{SMDfig1} but using the observed
SFH in HB06. From these figures, we find that both instantaneous
observed SFHs over-predict the SMH, and the SFH of MD14 gives a
better prediction. Moreover, we also find that the $f_r$ gives
different predictions comparing with a constant one especially at
high redshifts. From the middle panels of these two figures, we can
find that the SMHs predicted by observed SFHs with $f_r$ is little
higher than those with $R$, and the larger factors are up to about
$20\%$ at $z\sim8$ in both cases.

We also use the MCMC method to derive the best-fitted SFH from the
observed SMD data. We choose the SFH form as equation
(\ref{eq_MadauSFH}) used in MD14. And then we use equation
(\ref{relation_sfr_smd_2}) and MCMC method to fit the observed SMD
data and obtain the optimal parameters. Our result is
$(a,b,c,d) = (0.023,1.66,2.81,3.67)\pm(0.003,0.34,0.33,0.17)$, which
is quite different from the best-fitting parameters
$(a,b,c,d)=(0.015,2.7,2.9,5.6)$ obtained from observed SFR data
given in MD14. In Fig. \ref{SMDfig1}, the magenta solid line
represents the SMH predicted from our best-fitting SFH from observed
SMD data with SFH form in MD14 and the gray region shows the $95\%$
confidence region. We also use the SFH form in \cite{Cole2001}, and
obtain the optimal parameters $(a,b,c,d)=(0.030,0.058,2.361,2.707)$
$\pm$ (0.007,0.012,0.256,0.134) which is different with those of
HB06. \cite{Cole2001} obtained optimal parameters as
$(a,b,c,d)=(0.0166,0.1848,1.9474,2.6316)$ after considering the dust
extinction. The SFH is much lower if there is not dust absorption
and the optimal parameters are $(a,b,c,d)=(0.0,0.0798,1.658,3.105)$.
The difference of these two case is about factor of three at
$z\approx2$. Our best-fitting SFH from observed SMD data lies
between the SFHs of those two cases. Comparing with the predicted
SMHs from best-fitting SFHs from observed SMD data, we find that the
observed SFHs in both MD14 and HB06 over-predict the SMH. From the
bottom panel of Figs. \ref{SMDfig1}, we can find that the observed
SFH from MD14 over-predicts the SMD by a factor of about two at the
peak redshift about $z\sim1.5$. The bottom panel of \ref{SMDfig2}
shows that the observed SFH from HB06 over-predicts the SMD even by
a factor of about four at the peak redshift about $z\sim2$.

\begin{figure}[hbt]
  \centering
  \includegraphics[width=10cm]{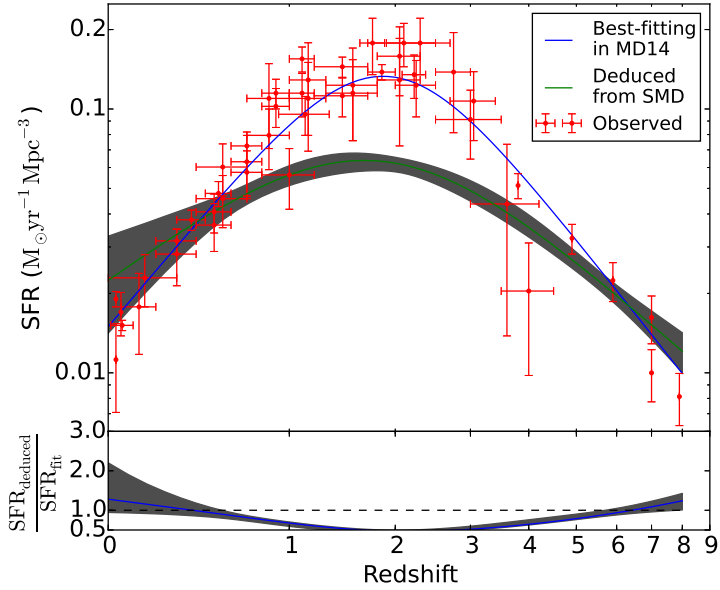}\\
  \caption{The SFH at $0<z<8$. Top panel: the red points with $1\sigma$ errors
  and the blue solid line are the observed SFR data and the best-fitting SFH given in MD14, respectively.
  The green solid line and gray region are our best-fitting SFH from observed SMD data and the $95\%$ confidence region obtained with MCMC method.
  The form of SFH using for MCMC fitting is given by MD14. Bottom panel: the ratio between our best-fitting SFH derived from observed SMD data and
  the best-fitting SFH in MD14.}\label{SFRfig1}
\end{figure}

From Fig. \ref{SFRfig1}, we find that the derived SFH from
observed SMD data is much different from that given in MD14.
Compared with the observed SFH of MD14, our best-fitting SFH is
consistent with it at $z<0.5$ and $z>6$. But in the range $0.5<z<6$,
our best-fitting SFH is lower. The lower factor peaks at about $z=2$
with about two. For the SFH form of \cite{Cole2001}, Fig.
\ref{SFRfig2} shows a similar result as Fig. \ref{SFRfig1}.
\cite{Wilkins2008} found that the observed SFH is consistent with
the SFH inferred from SMD at $z<1$ but about 4 times larger at
$z\approx3$, which is different with ours. It may be caused by the
different assumptions of IMF. Besides, \cite{Wilkins2008} only
considered the SFH and SMD at $z<4$. We consider a large redshift
range up to $z\sim 8$ and find that the SFH inferred from observed
SMD data is consistent with observed SFH at $z>6$.

\section{Discussion}
In this work, we use a large observational data sample to
test the discrepancy between the SMH and instantaneous SFH over the
redshift range $0<z<8$. We find that there is a discrepancy between
observed SMD and instantaneous SFH data. Just as said above, we
choose a single power-law Salpeter IMF and solar metallicity in our
analysis for simplicity. However, the estimations of SMD and SFRD
depend on the assumptions of IMF and metallicity. Therefore we
discuss the possible effect of the choice of IMF and metallicity on
our result and some other potential causes for this discrepancy in
this section.

\begin{figure}[hbt]
  \centering
  \includegraphics[width=8cm]{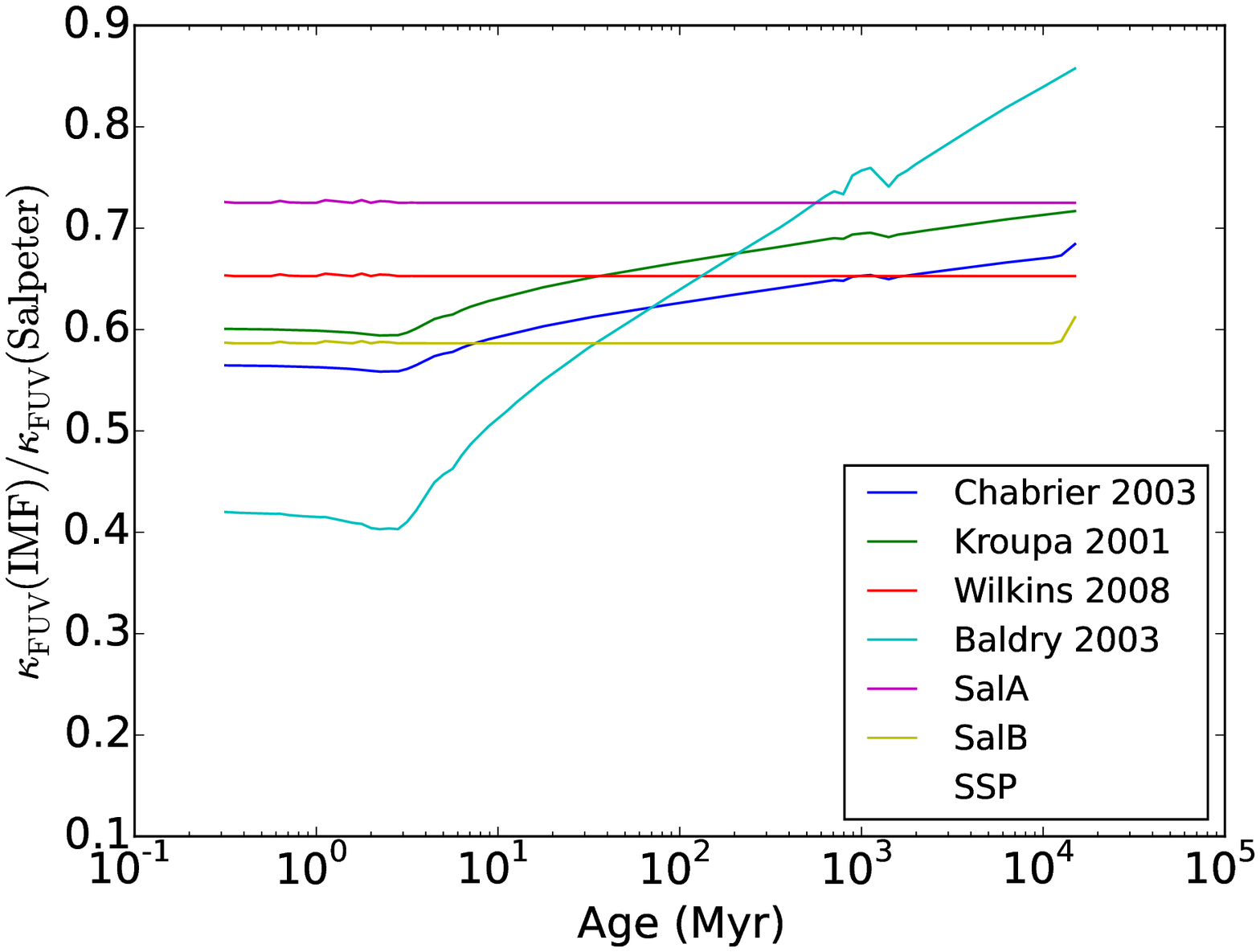}
  \includegraphics[width=8cm]{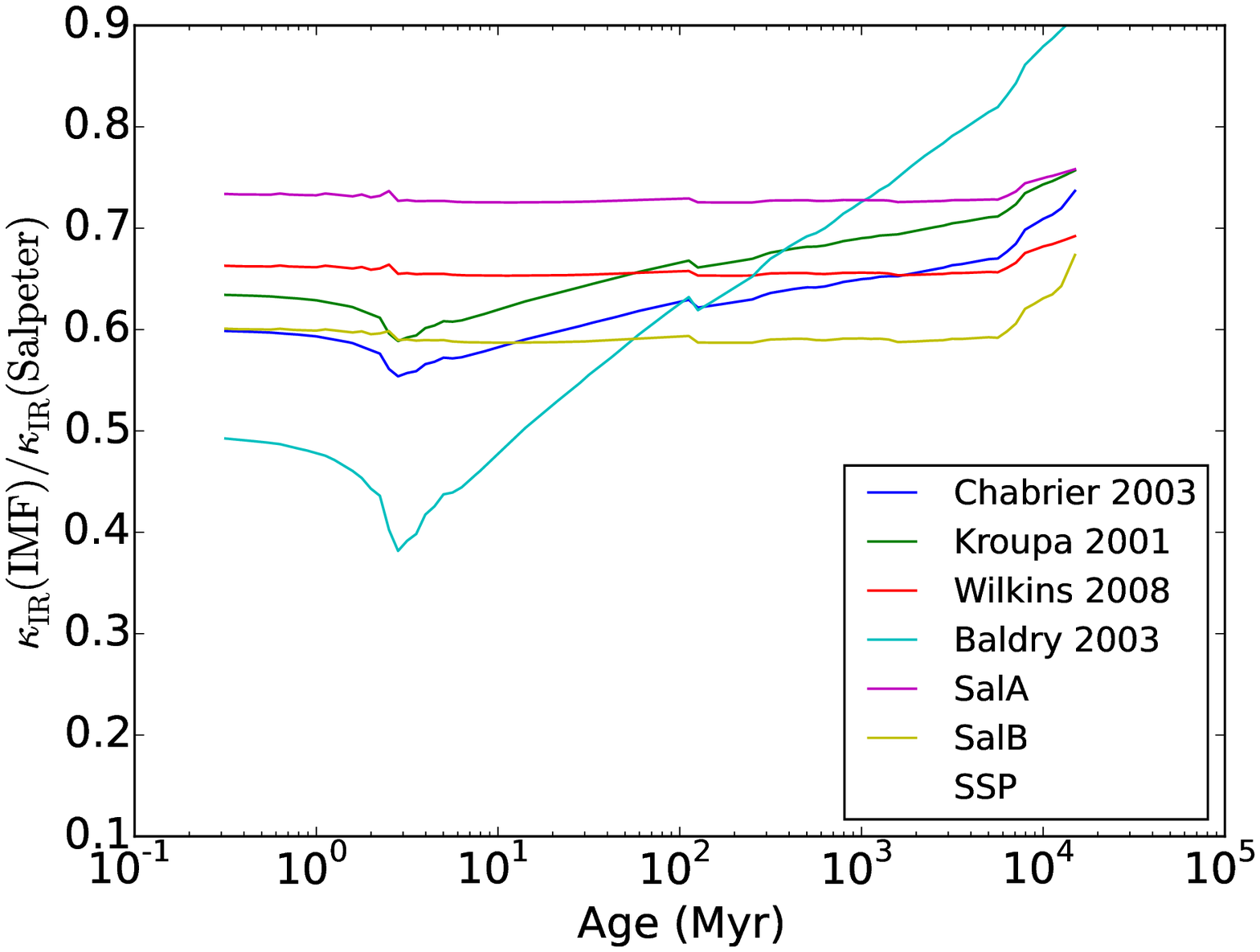}\\
  \includegraphics[width=8cm]{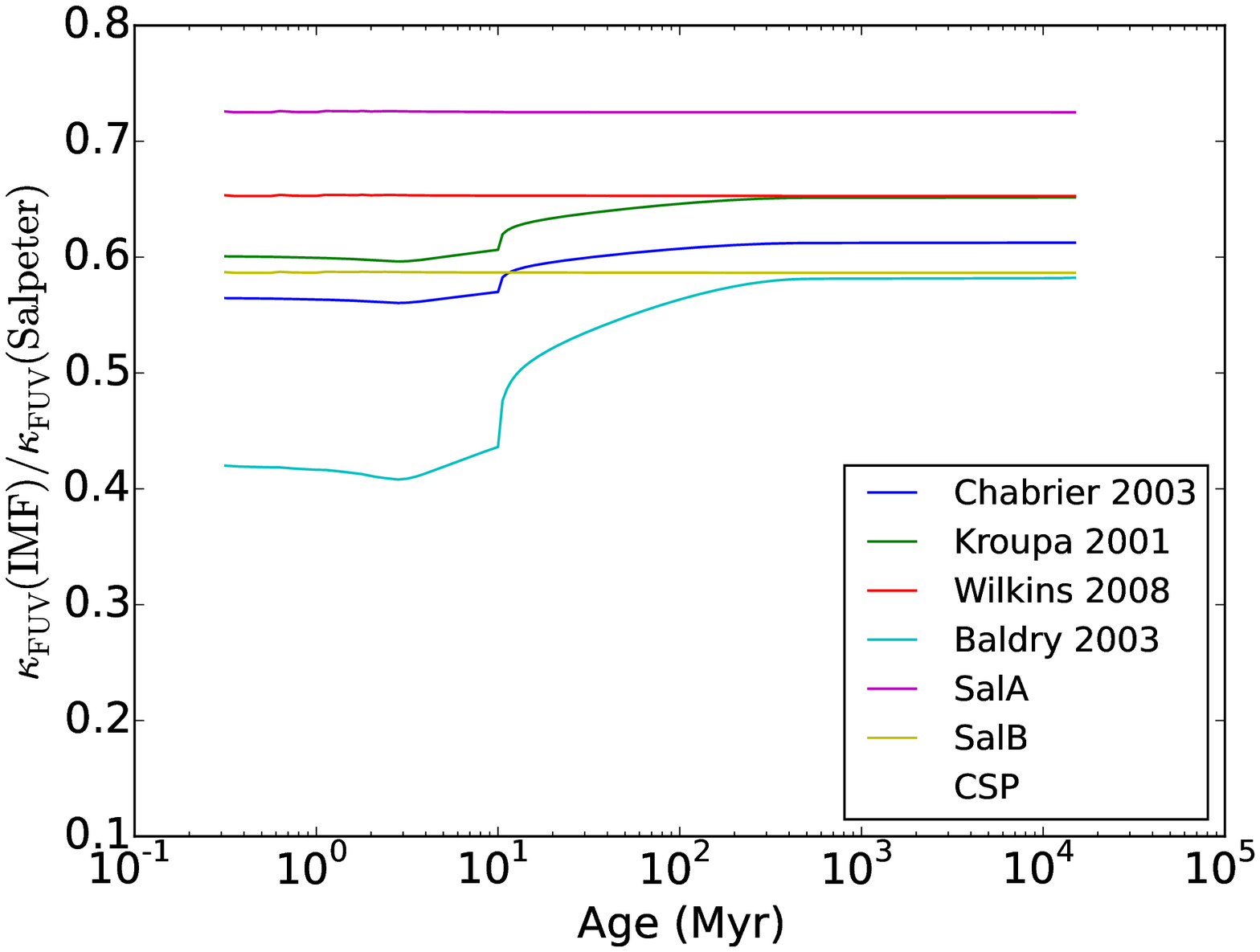}
  \includegraphics[width=8cm]{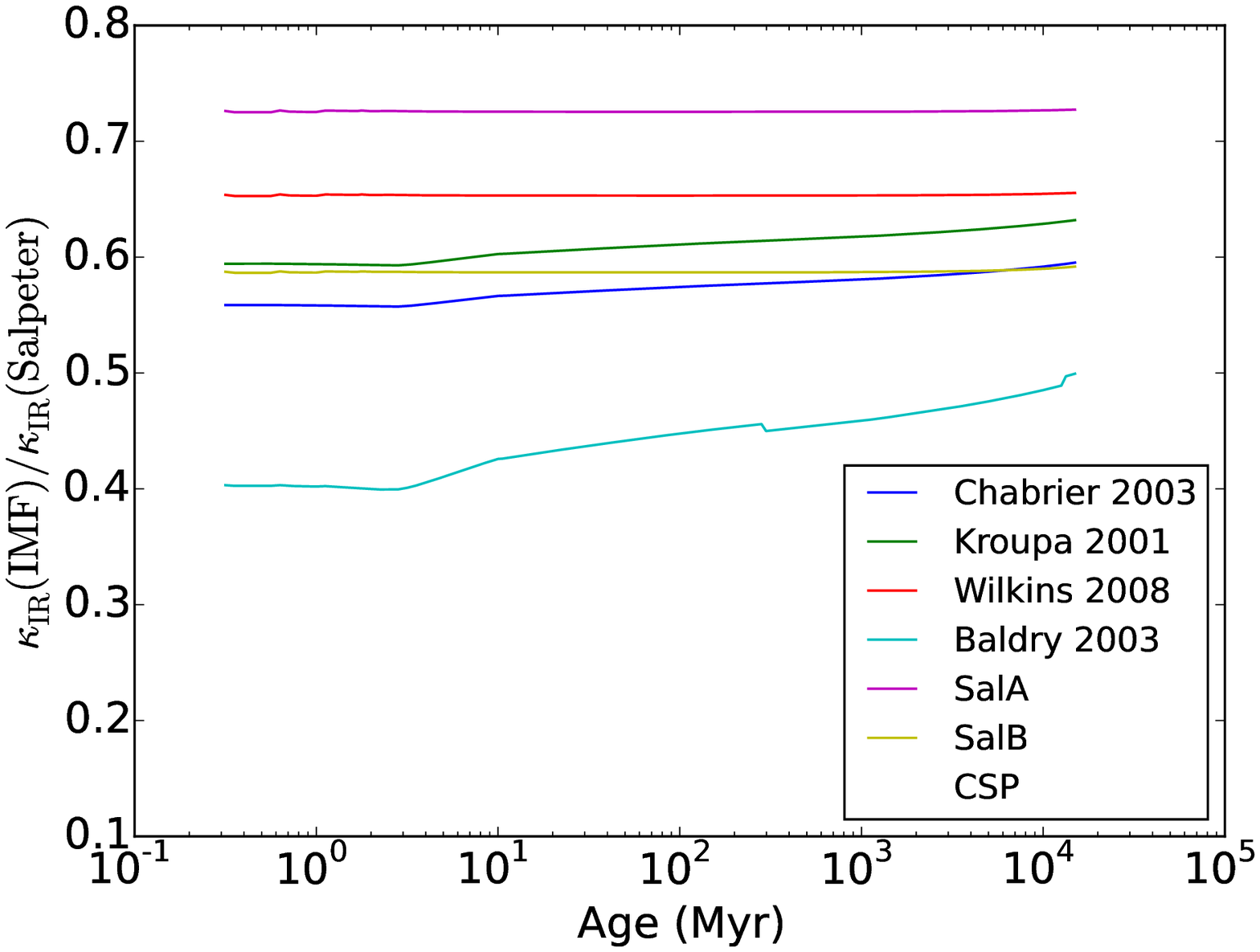}\\
  \caption{The ratio of mass-to-light from different IMFs to Salpeter IMF at FUV and IR bands.
  The top two panels is obtained from simple stellar populations, while the bottom two panels from
  complex stellar populations with a constant SFR.}\label{imfsLum}
\end{figure}

Generally, the high-mass slope of IMF affects the estimation
of SFRs since the young massive stars in a new born stellar
population dominant the emission especially for the UV emission.
While the low-mass slope of IMF affects the estimation of SMDs since
the old low-mass stars dominant the stellar mass of a galaxy.
Therefore, a more top-heavy IMF will generate more massive stars,
which emit more radiation leading to a high luminosity-to-SFR ratio.
Then we will obtain a low SFH for certain observed luminosity. Fig.
\ref{imfsLum} shows the converting factors of SFRD and SMD from
traditional Salpeter IMF to different IMFs. The top two panels,
which are obtained from simple stellar populations, show the
converting factors of SFRDs in UV and IR luminosity respectively.
The bottom two panels, which are obtained from complex stellar
populations with a constant SFR, are for the SMDs. For the
converting factors of SFRD, we choose the value in the first
$1\rm\,Myr$. While for factors of SMD we use the value after
$1\rm\,Gyr$ since the ratios become roughly unchanged in those time
region. The rough converting factors are listed in Table 1. These
converting factors are consistent with those given in previous
literature (HB06, W08, MD14).

\begin{figure}[hbt]
  \centering
  \includegraphics[width=10cm]{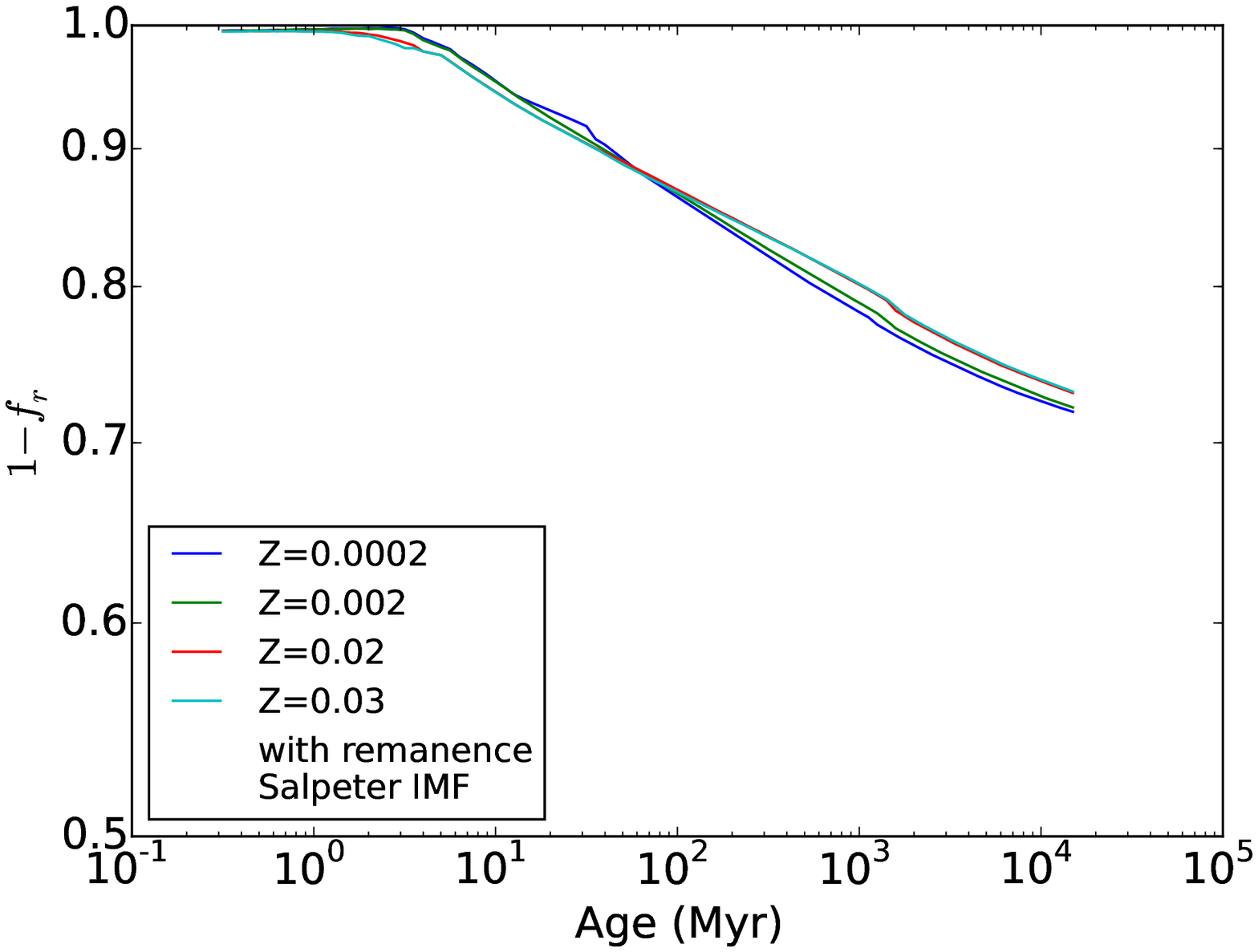}\\
  \includegraphics[width=10cm]{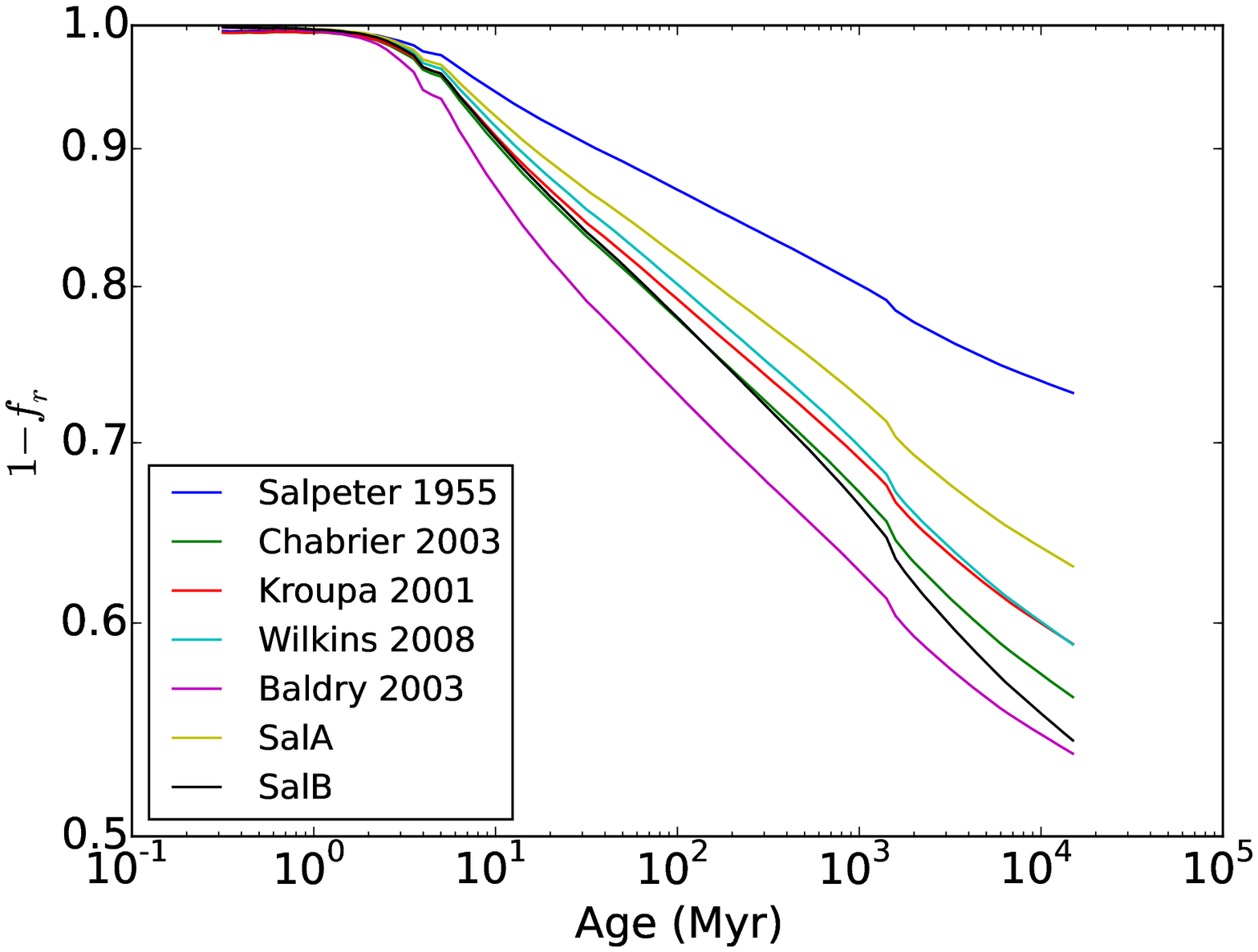}\\
  \caption{Top panel shows the dependence of $f_r$ on metallicity and the bottom one shows the dependence of $f_r$ on different IMFs.}\label{ftfig2}
\end{figure}

What's more, the choice of IMF can also affect the evolution
of the recycling factor of a stellar population, since the stars
with different initial mass have much different evolving processes
and mass-loss rates. The top panel in Fig. \ref{ftfig2} shows the
dependence of the recycling factor on different IMFs. We find that,
comparing with traditional Salpeter IMF, other IMFs with heavy
massive end or light low-mass end give larger recycling factors,
which can gives a lower predicted SMH from the observed SFH. The
rough recycling factors of different IMFs are listed in Table 1.
Considering the effects of the choice of different IMFs, the
discrepancy between the observed SFRD and SMD will be relieved if we
use the converting factors obtained with UV luminosity. For example,
if we consider the observed SFH given in MD14 and the converting factor of UV luminosity, the IMFs of
\cite{Chabrier2003} and \cite{Kroupa2001} will relieve the
discrepancy at about $40\%$ level, while the IMF of
\cite{Baldry2003} will relieve the discrepancy even at about $90\%$
level. The IMF of \cite{Baldry2003} can decrease the discrepancy at
about $30\%$ level even using the converting factors obtained from
IR luminosity.

Our current knowledge of the IMF remains remarkably poor
\citep{Kroupa2002}. Some observations suggest that the actual IMF
may deviate from the Salpeter IMF \citep{Dave2008,vanDokkum2008}.
The redshift distribution of gamma-ray bursts can be explained by an
evolving IMF \citep{Wang2011}. There are also some alternative IMFs
such as Kroupa IMF \cite{Kroupa2001} and Chabrier IMF
\cite{Chabrier2003}. However some observations show that the IMF in
local universe was not drastically different from Salpeter IMF over
a wide range of environments \citep{Weisz2015}. So, more accurate
and reliable observations and constraints on IMF are needed for the
following study of the cosmic SFH and SMH.

\begin{figure}[hbt]
  \centering
  \includegraphics[width=8cm]{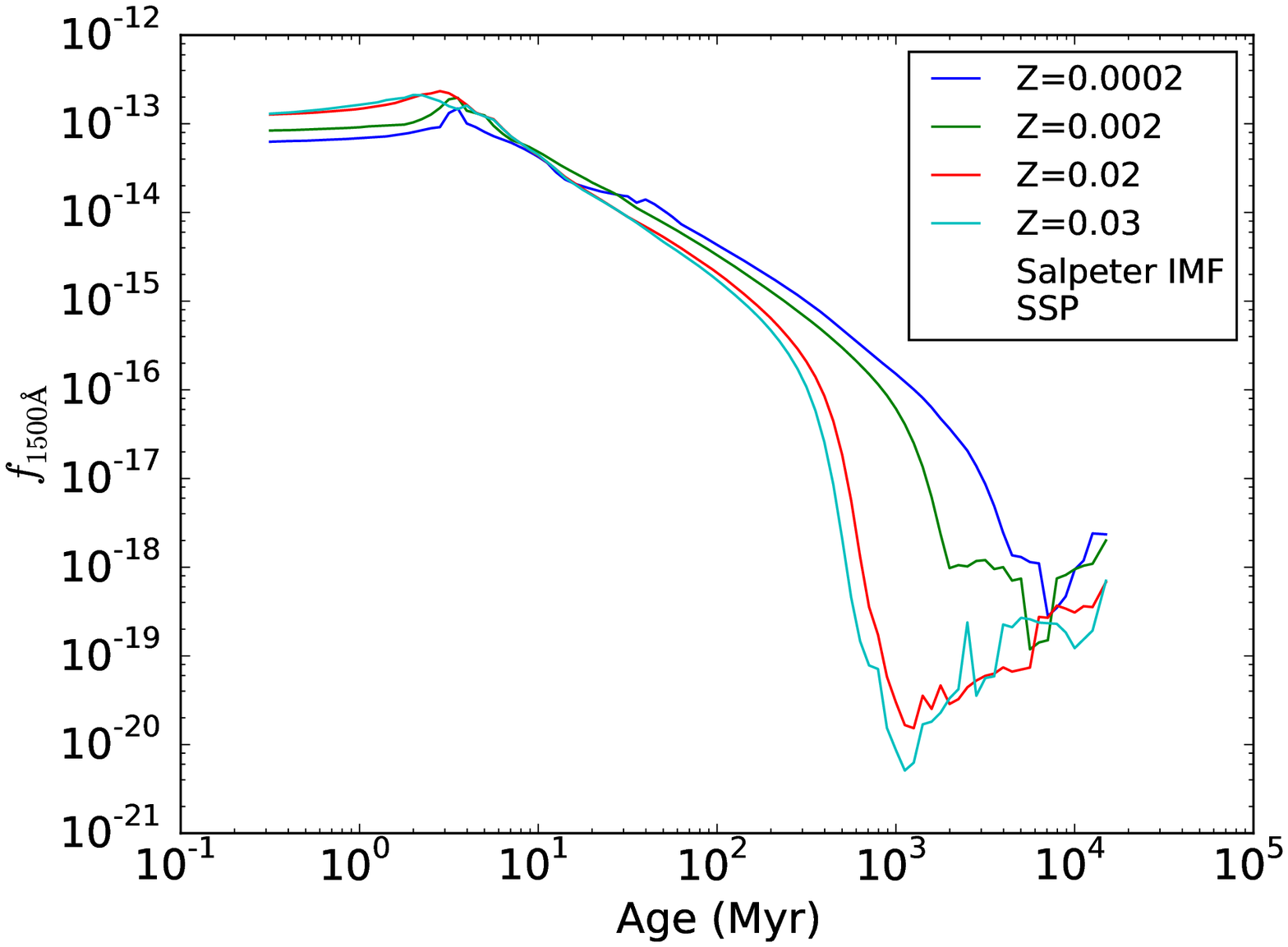}
  \includegraphics[width=8cm]{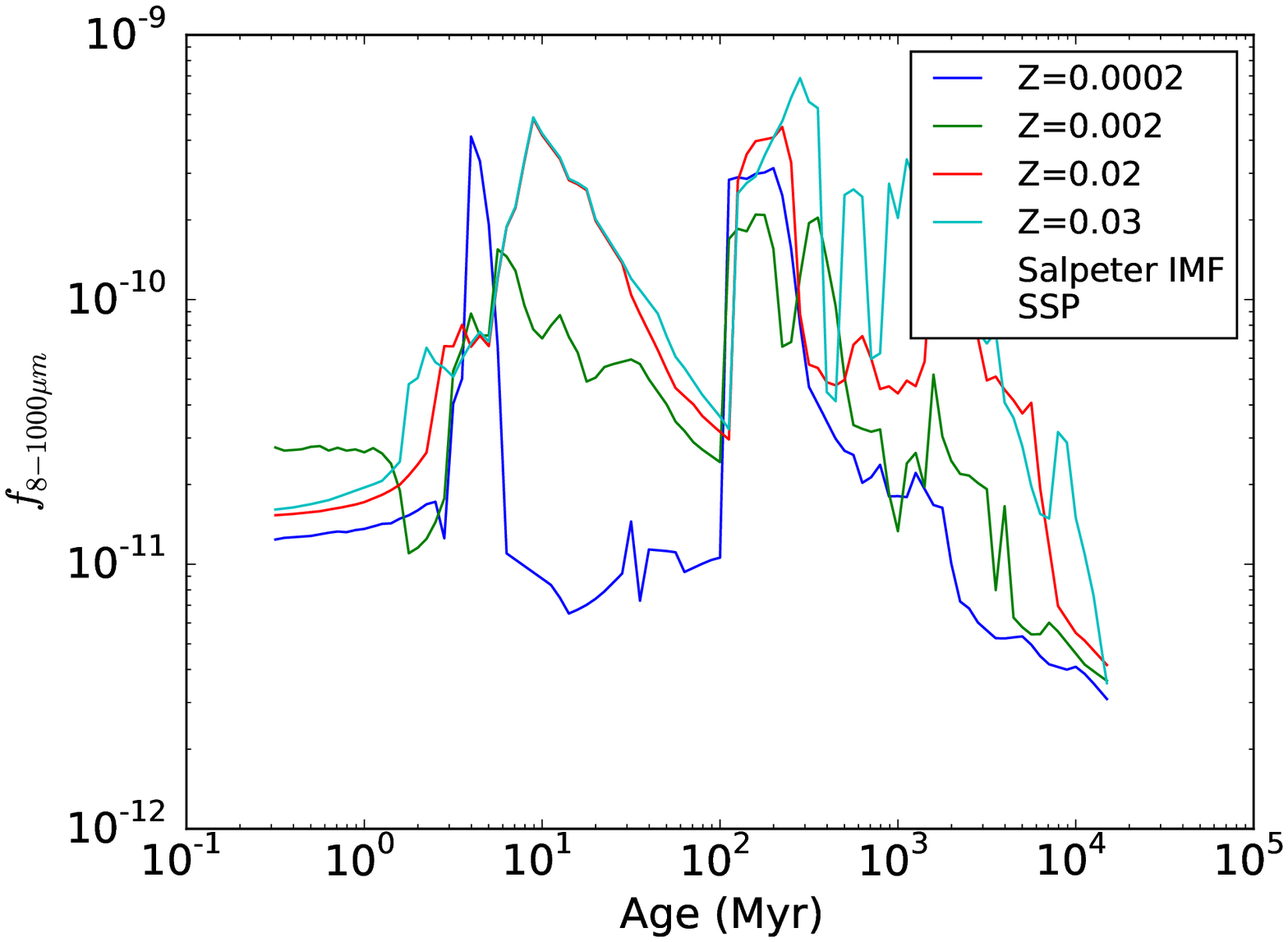}\\
  \includegraphics[width=8cm]{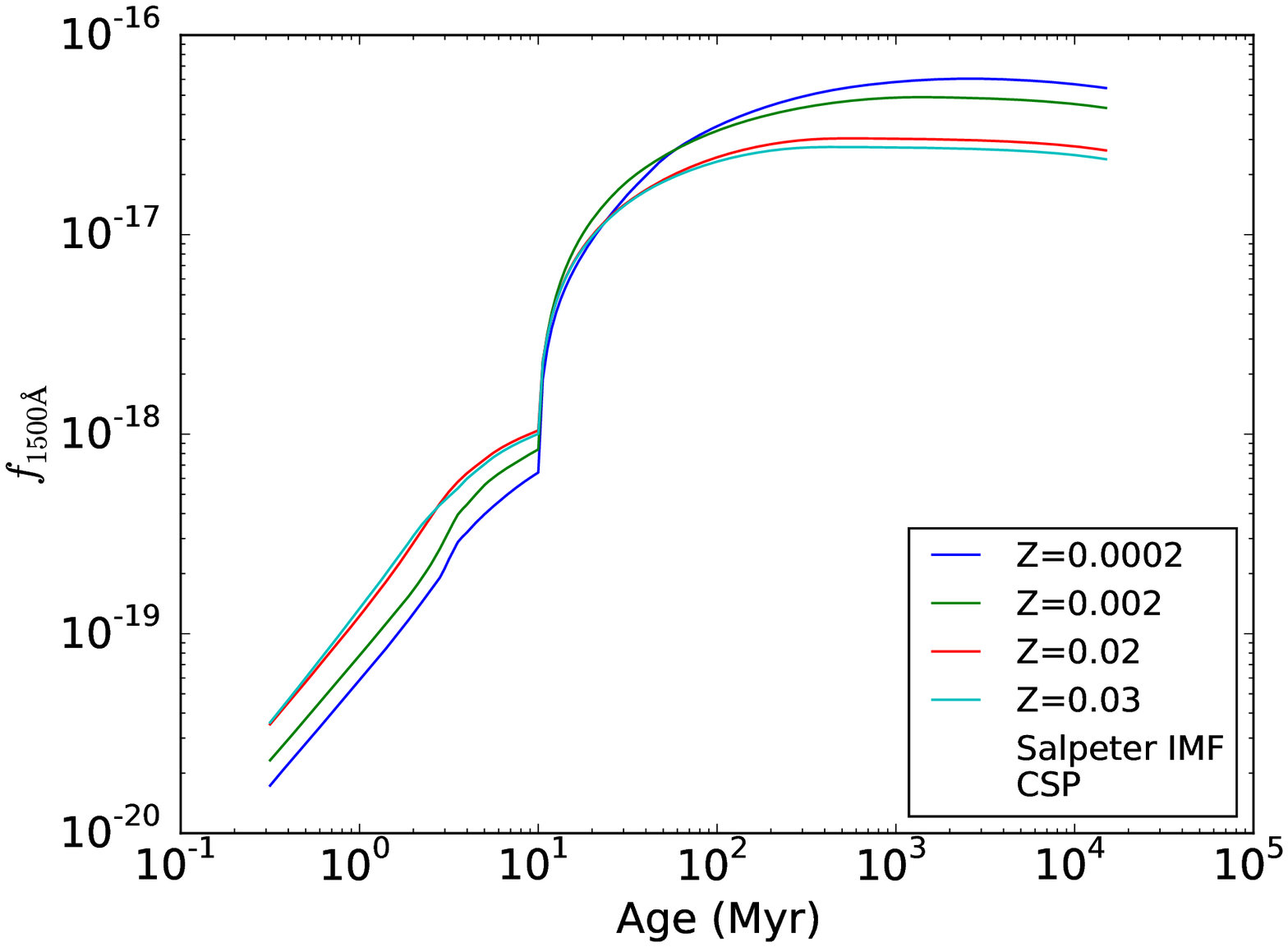}
  \includegraphics[width=8cm]{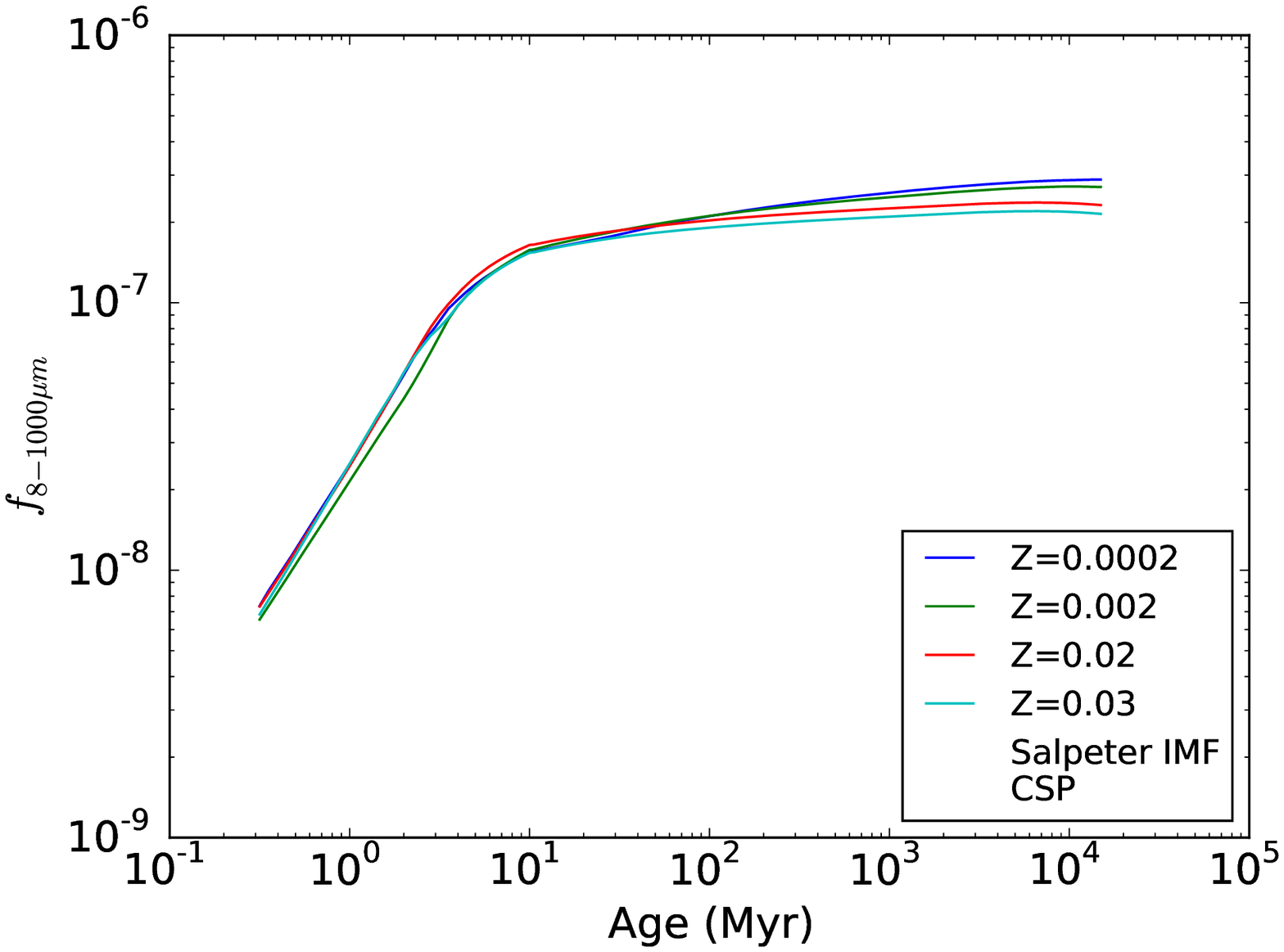}\\
  \caption{The dependence of UV and FIR luminosities on age and metallicity.
  The top two panels is obtained from simple stellar populations,
  while the bottom two panels from complex stellar populations with a constant SFR.}\label{metallicity}
\end{figure}

Except the IMF, the cosmic metallicity is also an important
factor for the estimation of SFH and SMH. The systematic effect of
the metallicity assumption needed to be discussed. We use the FSPS
code of \cite{Conroy2010} to calculate the evolution of $f_r$ and
the UV and IR luminosities of stellar populations under different
metallicities. The top panel in Fig. \ref{ftfig2} shows the
dependence of $f_r$ on metallicity in Salpeter IMF. We can see that
the dependence is not significant, since the $f_r$ only changes less
than $3\%$ while the metallicity changes 100 times. The top two
panels of Fig. \ref{metallicity} present the UV and IR luminosities
of different metallicities under the Salpeter IMF for the simple
stellar populations model. The bottom two panels for the complex
stellar populations model. From those figures, we can see that the
dependence of UV and IR luminosities on metallicity is significant
since it will change $1.5-4$ times while the metallicity changing
100 times. Considering the observations of cosmic metallicity
evolution, such as \cite{Rafelski2012} suggested the cosmic
metallicity will less than $0.1\,Z_\odot$ at $z>2$, this result
suggests that it would be better to take the effect of cosmic
metallicity evolution into account in further analysis.

From Figs. \ref{SMDfig1} and \ref{SMDfig2} we can find that the
observed SMD data are lower than the predicted value. Therefore, the
SMD might be systematically underestimated. As we known that the
mass of a galaxy is mainly dominated by its old low-mass stars but
luminosity is dominated by young massive stars which will lead to
the outshining problem. \cite{Maraston2010} studied the synthetic
spectrum of a composite population formed with a constant SFR for
$1~\rm{Gyr}$ and found the total spectrum was dominated by those
stars formed in the latest $0.5\rm~Gyr$. The stellar mass of those
galaxies with recent star formation will be systematically
underestimated since the older stars are lost in the bright light of
young stars. Therefore, we think that at the middle redshift range,
like $1<z<3$, the SFR is so large that it will lead to a
systematically underestimation of the SMDs since the outshining
problem.

\begin{figure}[hbt]
  \centering
  \includegraphics[width=10cm]{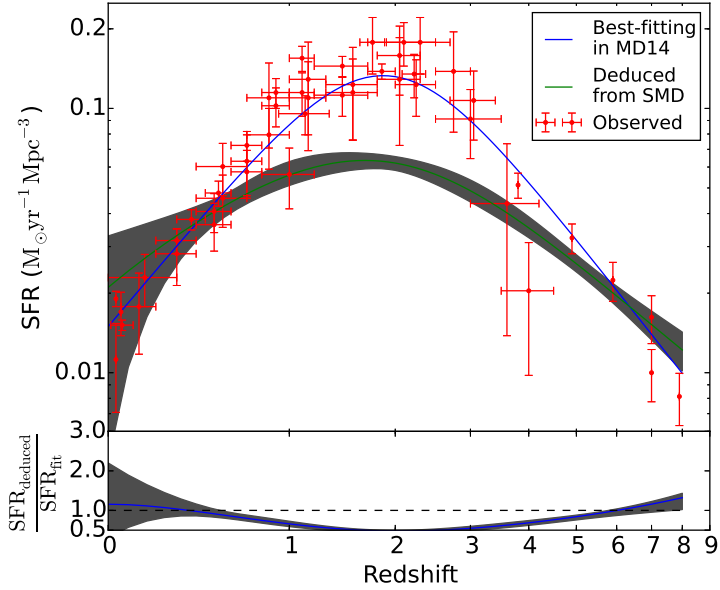}\\
  \caption{Same as Fig. \ref{SFRfig1} but the SFH form of \cite{Cole2001} is used for MCMC fitting.
  It shows a similar result as the Fig. \ref{SFRfig1}.}\label{SFRfig2}
\end{figure}

From Fig. \ref{SFRfig1} and Fig. \ref{SFRfig2} we can see that
although the observed SMD data are lower than the SMD inferred from
SFH at $z<0.5$. It doesn't mean that the best-fitting SFH are also
lower than the observed ones. In Fig. \ref{SFRfig1} and Fig.
\ref{SFRfig2}, the observed SFRs are much larger than the derived
SFRs at middle redshift range $0.5<z<6$. The observed SFR is about
two times as the derived one at $z\sim 2.0$. So the discrepancy
between SMD and SFH can be also caused by an overestimation of SFRs,
especially at middle redshift range $0.5<z<6$.
\cite{Cole2001} found that the SFH would be much lower if
there was not dust absorption. The difference of these two case is
about factor of three at $z\approx2$. Our best-fitting SFH from
observed SMD data lies between the SFHs of those two cases. This
hints that if we over-estimate the dust absorption effect, we will
over-estimate the observed SFRDs.

\section{Summary}
In this article, we use a large observational data sample to
test the discrepancy between the SMH and instantaneous SFH in the
redshift range $0<z<8$. At first, we integrate the observed SFH over
redshift to obtain the predicted SMH. We find that the inferred SMH
is overpredicted at redshift $z<4$ especially for the observed SFH
of H06. Although the SMH derived from the observed SFH of MD14 makes
a much better comparison, it still has overpredicted problem with a
factor of about two at redshift $z\sim1.5$. Comparing with the
results of the case using constant recycling factor $R$, the
evolving $f_r$ will give $20\%$ less SMH at $z\sim8$. Secondly, with
the form of SFH in MD14, we use MCMC method to fit the observed SMD
data and obtain the optimal parameters
$(a,b,c,d)=(0.028,1.88,2.40,3.69)\pm(0.005,0.47,0.36,0.29)$, which
are remarkably different with the result in MD14. The result is
shown in Fig.\ref{SMDfig1}. Fig.\ref{SFRfig1} shows the comparison
between the derived SFH from observed SMD data and the observed
SFRs. Comparing with the observed SFH of MD14, we can see that our
best-fitting SFH is consistent with it at $z<0.5$ and $z>6$. But in
the range of $0.5<z<6$, our best-fitting SFH is lower and even only half of observed
one at about $z=2$. In order to remove the possible effect of the
SFH form, we perform the same analysis with the form of SFH in
\cite{Cole2001}. For this SFH form, Fig. \ref{SFRfig2} shows a
similar result as Fig. \ref{SFRfig1}. \cite{Wilkins2008} found that
the observed SFH is consistent with the SFH inferred from SMD at
$z<1$ but about four times larger at $z\approx3$, which is different
with ours. Besides, \cite{Wilkins2008} only considered the SFH and
SMD at $z<4$, but we consider a large redshift range up to $z\sim 8$
and find that the SFH inferred from observed SMD data is consistent
with observed SFH at $z>6$.

We discuss some systematical effects of the assumptions of
IMF and metallicity. A top-heavy or bottom light IMF can relieve the
discrepancy between observed SFH and SMH. For example, if we
consider the observed SFH given in MD14, the IMFs of
\cite{Chabrier2003}, \cite{Kroupa2001} and \cite{Baldry2003} will
relieve the discrepancy at about $40\%$ to $90\%$ level,
respectively. We find that metallicity doesn't affect the evolution
of recycling factor $f_r$ significantly, which is only about $3\%$
with 100 times change in metallicity. However, the metallicity
affects the luminosities of UV and IR of stellar populations. We
also discuss the effect of possible under-estimation of SMD and
over-estimation of SFRD since the outshining problem and the
possible over-estimation of dust extinction.

In order to solve the discrepancy between the observed SMH
and SFH, we still need more accurate and reliable observations about
the evolution of cosmic IMF and metallicity. The next generation of
30-meter ground-based telescopes or space telescopes are expected.

\section*{Acknowledgments}
We thank an anonymous referee for useful suggestions and comments.
We also thank A. Grazian for providing some SMD data used in this
work. This work is supported by the National Basic Research Program
of China (973 Program, grant No. 2014CB845800), the National Natural
Science Foundation of China (grants 11422325 and 11373022), the
Excellent Youth Foundation of Jiangsu Province (BK20140016), and the
Program for New Century Excellent Talents in University (grant No.
NCET-13-0279).

\newpage

\newpage
\begin{table}\label{imfstable}
\caption{The power-law slopes, recycling factors of different IMFs,
and converting factors from Salpeter IMF to others IMFs.}
\begin{tabular}{lcccccc}
  \hline
  \multirow{2}{*}{IMF} & \multicolumn{3}{c}{Power-Law Slope (unit of mass: $M_\odot$)} & \multirow{2}{*}{$f_r^a$} & \multicolumn{2}{c}{Converting factor} \\
\cline{2-4}
\cline{6-7}
& $0.08-0.5$ & $0.5-1.0$ & $1.0-100$ & &SFRD$^b$& SMD \\ \hline
  Salpeter & 2.35 & 2.35 & 2.35 & 0.27 & 1 & 1 \\
  Chabrier & ...$^c$ & ... & -2.3 & 0.44 & 0.57 (0.59) & 0.61 (0.58) \\
  Kroupa & 1.3 & 2.3 & 2.3 & 0.41 & 0.60 (0.63) & 0.65 (0.62) \\
  Wilkins & 1.0 & 2.35 & 2.35 & 0.41 & 0.65 & 0.65 \\
  Baldry & 1.5 & 2.15 & 2.15 & 0.47 & 0.42 (0.48) & 0.58 (0.46) \\
  SalA & 1.5 & 2.35 & 2.35 & 0.37 & 0.73 & 0.73 \\
  SalB & 1.5 & 1.5 & 2.35 & 0.46 & 0.59 & 0.59 \\
  \hline
\end{tabular}
\begin{flushleft}
{
$^a$ The $f_r$ is measured at about 14 Gyr.\\
$^b$ For those IMFs with two converting factors, the converting factor in brackets are obtained by comparing the IR luminosity and the other is from UV luminosity.
For those IMFs with only one converting factor, it means the factors from UV and IR luminosities are same.\\
$^c$ The IMF of \cite{Chabrier2003} is $dn/dm\propto\exp{(-(\log{m}-\log{0.08})^2/(2\times0.69^2))}/m$ for $m\leq1\,M_\odot$.
}
\end{flushleft}
\end{table}

\newpage
\begin{deluxetable}{lcl}\label{SMDdata}
\tablecolumns{9} \tablewidth{0pc} \tabletypesize{\scriptsize}
\tablecaption{Stellar Mass Density Data.}
\tablehead{ \colhead{Redshift Range} &
\colhead{$\log(\rho_z^*)\rm~[M_{\odot}~Mpc^{-3}]$} & \colhead{Reference}} \startdata

  \multicolumn{3}{c}{Data listed below are from \cite{MadauDickinson2014} and} \\
  \multicolumn{3}{c}{represented by green dots in Fig. \ref{SMDfig1} and \ref{SMDfig2}} \\ \hline

  0.07 & $8.59_{-0.01}^{+0.01}$ & \cite{LiWhite2009} \\ \hline

  0.005-0.22 & $8.78_{-0.08}^{+0.07}$ & \cite{Gallazzi2008} \\ \hline

  0.0-0.2 & $8.59_{-0.05}^{+0.05}$ & \multirow{4}{*}{\cite{Moustakas2013}}\\
  0.2-0.3 & $8.56_{-0.09}^{+0.09}$ & \\
  0.3-0.4 & $8.59_{-0.06}^{+0.06}$ & \\
  0.4-0.5 & $8.55_{-0.08}^{+0.08}$ & \\ \hline

  0.2-0.4 & $8.46_{-0.12}^{+0.09}$ & \multirow{7}{*}{\cite{Bielby2012}} \\
  0.4-0.6 & $8.33_{-0.03}^{+0.03}$ & \\
  0.6-0.8 & $8.45_{-0.10}^{+0.08}$ & \\
  0.8-1.0 & $8.42_{-0.06}^{+0.05}$ & \\
  1.0-1.2 & $8.25_{-0.04}^{+0.04}$ & \\
  1.2-1.5 & $8.14_{-0.06}^{+0.06}$ & \\
  1.5-2.0 & $8.16_{-0.03}^{+0.32}$ & \\ \hline

  0.0-0.2 & $8.75_{-0.12}^{+0.12}$ & \multirow{12}{*}{\cite{Perez-Gonzalez2008}} \\
  0.2-0.4 & $8.61_{-0.06}^{+0.06}$ & \\
  0.4-0.6 & $8.57_{-0.04}^{+0.04}$ & \\
  0.6-0.8 & $8.52_{-0.05}^{+0.05}$ & \\
  0.8-1.0 & $8.44_{-0.05}^{+0.05}$ & \\
  1.0-1.3 & $8.35_{-0.05}^{+0.05}$ & \\
  1.3-1.6 & $8.18_{-0.07}^{+0.07}$ & \\
  1.6-2.0 & $8.02_{-0.07}^{+0.07}$ & \\
  2.0-2.5 & $7.87_{-0.09}^{+0.09}$ & \\
  2.5-3.0 & $7.76_{-0.18}^{+0.18}$ & \\
  3.0-3.5 & $7.63_{-0.14}^{+0.14}$ & \\
  3.5-4.0 & $7.49_{-0.13}^{+0.13}$ & \\ \hline

  0.2-0.5 & $8.55_{-0.09}^{+0.08}$ & \multirow{8}{*}{\cite{Ilbert2013}} \\
  0.5-0.8 & $8.47_{-0.08}^{+0.08}$ & \\
  0.8-1.1 & $8.50_{-0.08}^{+0.08}$ & \\
  1.1-1.5 & $8.34_{-0.07}^{+0.10}$ & \\
  1.5-2.0 & $8.11_{-0.06}^{+0.05}$ & \\
  2.0-2.5 & $7.87_{-0.09}^{+0.09}$ & \\
  2.5-3.0 & $7.64_{-0.14}^{+0.15}$ & \\
  3.0-4.0 & $7.24_{-0.20}^{+0.18}$ & \\ \hline

  0.2-0.5 & $8.61_{-0.06}^{+0.06}$ & \multirow{7}{*}{\cite{Muzzin2013}} \\
  0.5-1.0 & $8.46_{-0.03}^{+0.03}$ & \\
  1.0-1.5 & $8.22_{-0.03}^{+0.03}$ & \\
  1.5-2.0 & $7.99_{-0.03}^{+0.05}$ & \\
  2.0-2.5 & $7.63_{-0.04}^{+0.11}$ & \\
  2.5-3.0 & $7.52_{-0.09}^{+0.13}$ & \\
  3.0-4.0 & $6.84_{-0.20}^{+0.43}$ & \\ \hline

  0.3 & $8.78_{-0.16}^{+0.12}$ & \multirow{7}{*}{\cite{Arnouts2007}} \\
  0.5 & $8.64_{-0.11}^{+0.09}$ & \\
  0.7 & $8.62_{-0.10}^{+0.08}$ & \\
  0.9 & $8.70_{-0.15}^{+0.11}$ & \\
  1.1 & $8.51_{-0.11}^{+0.08}$ & \\
 1.35 & $8.39_{-0.13}^{+0.10}$ & \\
 1.75 & $8.13_{-0.13}^{+0.10}$ & \\ \hline

  0.1-0.35 & 8.58 & \multirow{4}{*}{\cite{Pozzetti2010}} \\
 0.35-0.55 & 8.49 & \\
 0.55-0.75 & 8.50 & \\
 0.75-1.00 & 8.42 & \\ \hline

  0.5-1.0 & 8.63 & \multirow{4}{*}{\cite{Kajisawa2009}} \\
  1.0-1.5 & 8.30 & \\
  1.5-2.5 & 8.04 & \\
  2.5-3.5 & 7.74 & \\ \hline

  1.3-2.0 & $8.11_{-0.02}^{+0.02}$ & \multirow{3}{*}{\cite{Marchesini2009}} \\
  2.0-3.0 & $7.75_{-0.04}^{+0.05}$ & \\
  3.0-4.0 & $7.47_{-0.13}^{+0.37}$ & \\ \hline

  1.9-2.7 & $8.10_{-0.03}^{+0.03}$ & \multirow{2}{*}{\cite{Reddy2012}} \\
  2.7-3.4 & $7.87_{-0.03}^{+0.03}$ & \\ \hline

  3.0-3.5 & $7.32_{-0.02}^{+0.04}$ & \multirow{3}{*}{\cite{Caputi2011}} \\
 3.5-4.25 & $7.05_{-0.10}^{+0.10}$ & \\
 4.25-5.0 & $6.37_{-0.54}^{+0.14}$ & \\ \hline

  3.8 & $7.24_{-0.06}^{+0.06}$ & \multirow{4}{*}{\cite{Gonzalez2011}} \\
  5.0 & $6.87_{-0.09}^{+0.08}$ & \\
  5.9 & $6.79_{-0.09}^{+0.09}$ & \\
  6.8 & $6.46_{-0.17}^{+0.14}$ & \\ \hline

  3.7 & $7.30_{-0.09}^{+0.07}$ & \multirow{2}{*}{\cite{Lee2012}} \\
  5.0 & $6.75_{-0.16}^{+0.33}$ & \\ \hline

  5.0 & $7.19_{-0.35}^{+0.19}$ & \cite{Yabe2009} \\ \hline

  8.0 & $5.78_{-0.30}^{+0.22}$ & \cite{Labbe2013} \\ \hline

  \multicolumn{3}{c}{Data listed below are used in \cite{Grazian2015} and} \\
  \multicolumn{3}{c}{represented by blue dots in Fig. \ref{SMDfig1} and \ref{SMDfig2}} \\ \hline

  3.5-4.5 & $7.36_{-0.03}^{+0.03}$ & \multirow{4}{*}{\cite{Grazian2015}} \\
  4.5-5.5 & $7.20_{-0.06}^{+0.04}$ & \\
  5.5-6.5 & $6.94_{-0.10}^{+0.10}$ & \\
  6.5-7.5 & $6.90_{-0.44}^{+0.34}$ & \\ \hline

  4.0 & $7.58_{-0.06}^{+0.06}$ & \multirow{4}{*}{\cite{Duncan2014}} \\
  5.0 & $7.39_{-0.07}^{+0.08}$ & \\
  6.0 & $6.98_{-0.11}^{+0.12}$ & \\
  7.0 & $6.86_{-0.56}^{+0.89}$ & \\ \hline

  0.625 & $8.42_{-0.08}^{+0.07}$ & \multirow{4}{*}{\cite{Tomczak2014}} \\
  0.875 & $8.36_{-0.13}^{+0.13}$ & \\
  1.125 & $8.24_{-0.12}^{+0.12}$ & \\
  1.375 & $8.09_{-0.08}^{+0.08}$ & \\
  1.75 & $8.09_{-0.08}^{+0.08}$ & \\
  2.25 & $7.94_{-0.17}^{+0.16}$ & \\ \hline

  0.6-1.0   & $ 8.45    _{- 0.06    }^{+    0.06    }$ &    \multirow{6}{*}{\cite{Santini2012}}\\
  1.0-1.4   & $ 8.28    _{- 0.08    }^{+    0.09    }$ &            \\
  1.4-1.8   & $ 8.16    _{- 0.07    }^{+    0.08    }$ &            \\
  1.8-2.5   & $ 8.18    _{- 0.11    }^{+    0.11    }$ &            \\
  2.5-3.5   & $ 8.10    _{- 0.13    }^{+    0.13    }$ &            \\
  3.5-4.5   & $ 7.91    _{- 0.24    }^{+    0.24    }$ &            \\ \hline
  6.95  & $ 6.57    _{- 0.14    }^{+    0.29    }$ &    \cite{Labbe2010}\\ \hline
  0.225 & $ 8.68    _{- 0.09    }^{+    0.09    }$ &    \multirow{6}{*}{\cite{Pozzetti2007}}\\
  0.55  & $ 8.57    _{- 0.08    }^{+    0.08    }$ &            \\
  0.8   & $ 8.45    _{- 0.11    }^{+    0.11    }$ &            \\
  1.05  & $ 8.37    _{- 0.12    }^{+    0.12    }$ &            \\
  1.4   & $ 8.27    _{- 0.14    }^{+    0.14    }$ &            \\
  2.05  & $ 8.28    _{- 0.11    }^{+    0.11    }$ &            \\ \hline
  0.95  & $ 8.46    _{- 0.07    }^{+    0.07    }$ &    \multirow{4}{*}{\cite{Dickinson2003}}\\
  1.70  & $ 8.06    _{- 0.17    }^{+    0.13    }$ &            \\
  2.25  & $ 7.58    _{- 0.11    }^{+    0.07    }$ &            \\
  2.75  & $ 7.52    _{- 0.15    }^{+    0.14    }$ &            \\ \hline
  0.3   & $ 8.68    _{- 0.05    }^{+    0.06    }$ &    \multirow{7}{*}{\cite{Ilbert2010}}\\
  0.5   & $ 8.44    _{- 0.04    }^{+    0.05    }$ &            \\
  0.7   & $ 8.44    _{- 0.03    }^{+    0.03    }$ &            \\
  0.9   & $ 8.50    _{- 0.03    }^{+    0.03    }$ &            \\
  1.1   & $ 8.28    _{- 0.04    }^{+    0.03    }$ &            \\
  1.35  & $ 8.13    _{- 0.03    }^{+    0.03    }$ &            \\
  1.75  & $ 7.97    _{- 0.24    }^{+    0.16    }$ &            \\ \hline
  3.52  & $ 7.05    _{- 0.77    }^{+    0.76    }$ &    \cite{Marchesini2010}\\ \hline
  1.25  & $ 8.35    _{- 0.11    }^{+    0.10    }$ &    \multirow{5}{*}{\cite{Mortlock2011}}\\
  1.75  & $ 8.36    _{- 0.12    }^{+    0.10    }$ &            \\
  2.25  & $ 8.23    _{- 0.14    }^{+    0.12    }$ &            \\
  2.75  & $ 8.21    _{- 0.11    }^{+    0.12    }$ &            \\
  3.25  & $ 7.87    _{- 0.11    }^{+    0.10    }$ &            \\ \hline
  0.5   & $ 8.46    _{- 0.03    }^{+    0.03    }$ &    \multirow{8}{*}{\cite{Fontana2006}}\\
  0.7   & $ 8.37    _{- 0.02    }^{+    0.02    }$ &            \\
  0.9   & $ 8.29    _{- 0.03    }^{+    0.03    }$ &            \\
  1.15  & $ 8.18    _{- 0.02    }^{+    0.02    }$ &            \\
  1.45  & $ 8.07    _{- 0.03    }^{+    0.03    }$ &            \\
  1.8   & $ 7.94    _{- 0.04    }^{+    0.04    }$ &            \\
  2.5   & $ 7.68    _{- 0.04    }^{+    0.04    }$ &            \\
  3.5   & $ 7.27    _{- 0.02    }^{+    0.03    }$ &            \\ \hline

\enddata
\\
\end{deluxetable}

\end{document}